\documentclass[10pt]{elsart3p}

\usepackage{natbib}
\usepackage{graphics}
\usepackage{amssymb}
\usepackage{epsfig}
\usepackage[dvips]{color}
\newcommand{\ber}{\mbox{${^7}$Be}}
\newcommand{\che}{\v{C}erenkov~}
\newcommand{\Bipo}{\mbox{${^{214}}$Bi-${^{214}}$Po}}
\newcommand{\bipo}{\mbox{${^{212}}$Bi-${^{212}}$Po}}

\begin{document}

\begin{frontmatter}

\title{First real time detection of $^{7}$Be solar neutrinos by Borexino}
\vspace{5mm}
\subtitle{\large Borexino Collaboration}

\author[LNGS]{C.~Arpesella\thanksref{label}},
\thanks[label]{deceased}
\author[Milano]{G.~Bellini},
\author[Princeton2]{J.~Benziger},
\author[Milano]{S.~Bonetti},
\author[Milano]{B.~Caccianiga},
\author[Princeton1]{F.~Calaprice}, 
\author[Princeton1]{F.~Dalnoki-Veress},
\author[Milano]{D.~D'Angelo},
\author[APC]{H.~de Kerret},
\author[Peters]{A.~Derbin},
\author[MIT]{M.~Deutsch\thanksref{label}},
\author[Kurchatov]{A.~Etenko},
\author[Dubna]{K.~Fomenko},
\author[Princeton1]{R.~Ford\thanksref{label3}},
\thanks[label3]{now at SNOLAB, Sudbury, Canada}
\author[Milano]{D.~Franco},
\author[Heidelberg]{B.~Freudiger\thanksref{label}},
\author[Princeton1]{C.~Galbiati},
\author[LNGS]{S.~Gazzana},
\author[Milano]{M.~Giammarchi},
\author[Munich]{M.~Goeger-Neff},
\author[Princeton1]{A.~Goretti},
\author[Virginia]{C.~Grieb},
\author[Virginia]{S.~Hardy},
\author[Heidelberg]{G.~Heusser},
\author[LNGS]{Aldo~Ianni},
\author[Princeton1]{Andrea~Ianni},
\author[Virginia]{M.~Joyce},
\author[LNGS]{G.~Korga},
\author[APC]{D.~Kryn},
\author[LNGS]{M.~Laubenstein},
\author[Princeton1]{M.~Leung},
\author[Kurchatov]{E.~Litvinovich},
\author[Milano]{P.~Lombardi},
\author[Milano]{L.~Ludhova},
\author[Kurchatov]{I.~Machulin},
\author[Genova]{G.~Manuzio},
\author[Kurchatov]{A.~Martemianov\thanksref{label}},
\author[Perugia]{F.~Masetti},
\author[Princeton1]{K.~McCarty},
\author[Milano]{E.~Meroni},
\author[Milano]{L.~Miramonti},
\author[Cracow]{M.~Misiaszek},
\author[LNGS]{D.~Montanari},
\author[LNGS]{M.~E.~Monzani},
\author[Peters]{V.~Muratova},
\author[Munich]{L.~Niedermeier},
\author[Munich]{L.~Oberauer},
\author[APC]{M.~Obolensky},
\author[Perugia]{F.~Ortica},
\author[Genova]{M.~Pallavicini}, 
\author[LNGS]{L.~Papp},
\author[Milano]{L.~Perasso},
\author[Princeton1]{A.~Pocar\thanksref{label2}},
\thanks[label2]{now at Department of Physics, 
Stanford University, Stanford, CA, USA}
\author[Virginia]{R.~S.~Raghavan},
\author[Milano]{G.~Ranucci} ,
\author[LNGS]{A.~Razeto} ,
\author[Kurchatov]{A.~Sabelnikov},
\author[LNGS]{C.~Salvo},
\author[Heidelberg]{S.~Sch\"onert} ,
\author[Heidelberg]{H.~Simgen} ,
\author[Dubna]{O.~Smirnov},
\author[Kurchatov]{M.~Skorokhvatov},
\author[Princeton1]{A.~Sonnenschein\thanksref{label4}},
\thanks[label4]{now at Fermi National Accelerator Laboratory, Batavia, IL, USA}
\author[Dubna]{A.~Sotnikov},
\author[Kurchatov]{S.~Sukhotin},
\author[Milano,Kurchatov]{Y.~Suvorov} ,
\author[Kurchatov]{V.~Tarasenkov},
\author[LNGS]{R.~Tartaglia} ,
\author[Genova]{G.~Testera} ,
\author[APC]{D.~Vignaud},
\author[Genova]{S.~Vitale\thanksref{label}},
\author[Virginia]{R.~B.~Vogelaar} ,
\author[Munich]{F.~von~Feilitzsch} ,
\author[Cracow]{M.~Wojcik},
\author[Dubna]{O.~Zaimidoroga},
\author[Genova]{S.~Zavatarelli} ,
\author[Heidelberg]{G.~Zuzel}

\address[LNGS]{I.N.F.N. Laboratori Nazionali del Gran Sasso -- Assergi -- Italy}
\address[Milano]{Dipartimento di Fisica dell'Universit\`a degli Studi e INFN -- Milano -- Italy}
\address[Princeton2]{Princeton University, Chemical Engineering Department -- Princeton, NJ -- USA}
\address[Princeton1]{Princeton University, Physics Department -- Princeton, NJ -- USA}
\address[Kurchatov]{RRC Kurchatov Institute -- Moscow -- Russia}
\address[APC]{Laboratoire AstroParticule et Cosmologie APC -- Paris -- France}
\address[Peters]{St. Petersburg Nuclear Physics Institute -- Gatchina -- Russia}
\address[MIT]{Massachussetts Institute of Technology, Department of Physics -- Cambridge, MA -- USA}
\address[Dubna]{Joint Institute for Nuclear Research -- Dubna -- Russia}
\address[Munich]{Technische Universit\"at Muenchen -- Garching -- Germany}
\address[Virginia]{Virginia Tech,  Physics Department -- Blacksburg, VA -- USA}
\address[Genova]{Dipartimento di Fisica dell'Universit\`a e INFN -- Genova -- Italy}
\address[Perugia]{Dipartimento di Chimica dell'Universit\`a e INFN -- Perugia -- Italy}
\address[Cracow]{Marian Smoluchowski Institute of Physics, Jagiellonian University -- Krakow -- Poland}
\address[Heidelberg]{Max-Planck-Institut f\"ur Kernphysik -- Heidelberg -- Germany}

\begin{abstract}
This paper reports a direct measurement of the \ber\ solar neutrino signal rate performed with the Borexino low background liquid scintillator detector.  This is the first real-time spectral measurement of sub-MeV solar neutrinos.  The result for 0.862~MeV \ber\ neutrinos is $47\pm7_{\rm stat}\pm12_{\rm sys}$~counts/(day $\cdot$ 100~ton), consistent with predictions of Standard Solar Models and neutrino oscillations with LMA-MSW parameters.
\end{abstract}


\end{frontmatter}

\section{Introduction}
\label{sec:intro}

We report a real-time spectroscopic observation of the mono-energetic (0.862~MeV) neutrinos ($\nu$) from the radioactive decay of~\ber\ in the Sun.  These results from Borexino at the Laboratori Nazionali del Gran Sasso constitute a scientific and technological breakthrough in the quest for the real time detection of low energy (below 2~MeV) solar $\nu$ that comprise $>$99\%~of the flux.  Current spectroscopic data on solar $\nu$ are confined to energies greater than 5~MeV (0.01\%~of the total flux).

Solar neutrinos provide a unique probe for studying both the nuclear fusion reactions that power the Sun and the fundamental properties of neutrinos.

The sun shines by the energy produced in a thermonuclear reaction chain built up on the fusion of four protons to produce one $^{4}$He nucleus.  The chain has three major links: {\it pp}I -- the fundamental proton-proton fusion leading to the production of $^{3}$He; {\it pp}II -- the production of~\ber\ via the reaction $^{3}$He+$^{4}$He; {\it pp}III -- the production of $^{8}$B via $^{7}$Be+$p$, each step terminating with the production and accumulation of $^{4}$He.  Each of these links produces electron neutrinos ($\nu_e$).  The~\ber\ from the {\it pp}II link decays by electron capture with a $\simeq$90\% branch yielding mono-energetic 0.862 MeV neutrinos.  The~\ber~$\nu_e$ contribute a substantial ($\simeq$10\%) fraction of the solar $\nu_e$ flux, whereas only $10^{-4}$ of the total $\nu_e$ flux arises from {\it pp}III.  The $^{8}$B neutrinos from {\it pp}III have been intensively studied in the last decade~\cite{bib:sk,bib:sno}.

The prediction of the \ber~$\nu$ flux depends both on the solar model and on the cross section of the $^{3}$He($\alpha,\gamma$)\ber\ reaction.  The latter was recently measured with improved accuracy~\cite{bib:luna}; therefore a measurement of the \ber~$\nu$ flux is an important and timely test of the standard solar model.

Precision measurements in Borexino are equally important for the new neutrino physics with non-zero masses, flavor mixing, and neutrino oscillations.  Current neutrino data are consistent with the so-called Large Mixing Angle (LMA) solution of the Mikheyev-Smirnov-Wolfenstein (MSW) picture of flavor conversion~\cite{bib:bahcall,bib:fogli}.  The LMA solution predicts a transition from matter enhanced oscillations at $^{8}$B energies to vacuum oscillations at low energies in the neighborhood of \ber~neutrinos.  The $\nu_e$ survival probability increases from 0.33 at high energies up to $\approx$0.6 at low energies near the \ber~$\nu_e$ line.  A measurement of the \ber~neutrino rate will test the predicted increase in the $\nu_e$ survival probability.  New physics such as non-standard interactions 
introduce deviations from these expectations~\cite{bib:friedland}.

Solar neutrinos have been detected in the last 40 years by two methods, the earliest being radiochemical separation after neutrino activation which yields only the integrated $\nu_e$ flux above a threshold. Individual components of the neutrino spectrum cannot be determined by such measurements.  Low energy neutrinos have been observed so far only by such methods~\cite{bib:homestake,bib:gallex,bib:sage,bib:gno}.  Direct spectroscopy using kiloton scale water \che detectors has been limited to the high energy neutrino flux above 5 MeV, owing to the low signal light yield of the \che process and to the natural radioactivity background~\cite{bib:sk,bib:sno}.

\section{Experimental apparatus and signature from solar neutrinos}
\label{sec:detector}

\begin{center}
\begin{figure}
\centering\epsfig{file=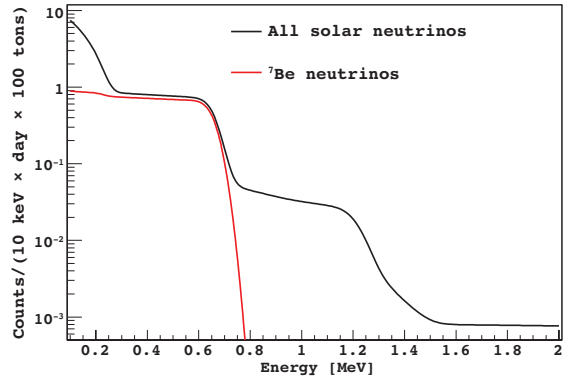, width=0.50\textwidth}
\caption{Neutrino spectra expected in Borexino (accounting for the detector's energy resolution).  The solid black line represents the neutrino signal rate in Borexino according to the most recent predictions of the Standard Solar Model~\cite{bib:carlos} including neutrino oscillations with the LMA-MSW parameters.  The solid red line illustrates the contribution due to \ber~neutrinos.  {\it pp} neutrinos contribute to the spectrum below 0.3~MeV and the edge at 1.2~MeV is due to {\it pep} neutrinos.}
\label{fig:teoflux}
\end{figure}
\end{center}

Borexino employs a liquid scintillator that produces sufficient light to observe low energy neutrino events via elastic scattering by electrons.  The reaction is sensitive to all neutrino flavors by the neutral current interaction, but the cross section for $\nu_e$ is larger due to the combination of charged and neutral currents.  The recoil electron profile for a mono-energetic neutrino is similar to that of Compton scattering of a single $\gamma$-ray.  Thus, the recoil electron profile is basically a rectangular shape with a sharp cut-off edge at 665~keV in the case of \ber~neutrinos (see Fig.~\ref{fig:teoflux}).  The background from the 156~keV $\beta$~decay of $^{14}$C, intrinsic to the scintillator, limits neutrino observation to energies above 200~keV.

Organic liquid scintillation technology is a viable method for massive detectors suitable for spectroscopy at such low energies, since its signal efficiency is $\approx$50 ~times that of the commonly used \che technology.  However, no directionality is possible.  Moreover, it is not possible to distinguish neutrino scattered electrons from electrons due to natural radioactivity.  Thus the key requirement in the technology of Borexino is extremely low radioactive contamination.

The scintillator is pseudocumene (PC, 1,2,4-trimethylbenzene), a well known aromatic scintillation solvent doped with   PPO (2,5-diphenyloxazole, a fluorescent dye) at a concentration of 1.5~g/l.  The excitation energy of PC molecules produced by ionizing radiation is transferred non-radiatively to the PPO which then radiates with a decay time of 1.3~ns resulting in a scintillation pulse width of 3.5~ns.  The scintillation signal is emitted in the optical band centered near 400~nm. At this wavelength the optical attenuation length in the scintillator is~$\simeq$7~m.

\begin{center}
\begin{figure}
\centering\epsfig{file=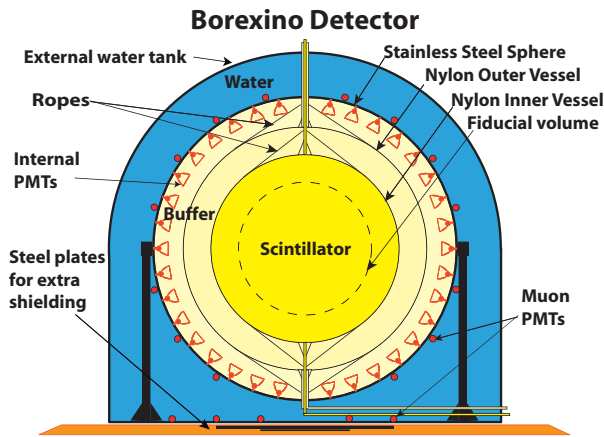, width=0.50\textwidth}
\caption{Schematic drawing of the Borexino detector.}
\label{fig:detector}
\end{figure}
\end{center}

The PC was produced by Polimeri Europa in Sardinia, Italy. Precautions, including specially constructed loading and unloading stations, were taken to minimize the optical and radioactive contamination during the transportation to Gran Sasso.  In particular, the PC was taken directly from the production plant and quickly shipped in a specially constructed nitrogen pressurized vessel.

Techniques for purification of the scintillator employed in this work are based mainly on methods developed and tested in earlier studies with the Counting Test Facility (CTF), a 4-ton prototype of Borexino.  CTF demonstrated for the first time the feasibility of achieving the low backgrounds needed to detect solar neutrinos in a large scale scintillator~\cite{bib:ctf1,bib:ctf2,bib:ctf3}.  For Borexino a larger purification plant was developed similar to the CTF system,  but with several improved features including the use of high vacuum and precision cleaning techniques.  To achieve ultra low operating backgrounds in the detector (from internal and external sources) the design of Borexino is based on the principle of graded shielding with the scintillator at the center of a set of concentric shells of increasing radiopurity (see Fig.~\ref{fig:detector}).  The 300-ton scintillator is contained in a thin (125~$\mu$m) nylon Inner Vessel (IV) with a radius of 4.25~m~\cite{bib:nylon}.  Within the IV a fiducial mass is defined by software selection of the events based on their reconstructed position using timing data from the PMTs.

A second nylon outer vessel (OV) with radius 5.50~m contains a passive shield composed of pseudocumene and 5.0~g/l DMP (dimethylphthalate), a material that quenches the residual scintillation of PC so that spectroscopic signals arise dominantly from the interior of the IV.  The OV acts as a barrier against radon and other background contaminations originating from outside. The third vessel is a stainless steel sphere (SSS) with radius 6.85~m that encloses the PC-DMP buffer fluid which fills the space between the OV and the SSS. The SSS also serves as a support structure for the PMTs.  Finally, the entire detector is contained in a tank (radius 9~m, height 16.9~m) of ultra-pure water. The total liquid passive shielding of the central volume from external radiation (such as the rock) is thus~$\simeq$5.5~m of water equivalent (m.w.e).  The scintillator material in the IV is less dense than the buffer fluids by about~0.1\%; this results in a slight upward buoyancy force on the IV.  Thin low-background ropes made of ultra-high density polyethylene hold the nylon vessels in place.

The scintillation light is viewed by 2212 8'' PMTs  (ETL 9351) uniformly distributed on the inner surface of the SSS~\cite{bib:pmts1,bib:pmts2}.  All but~384~photomultipliers are equipped with aluminum light concentrators designed to increase the collection efficiency of the light from the scintillator and to reduce the number of photons not coming from the scintillator volume~\cite{bib:cones}.  Residual background scintillation and \che light that escape quenching in the buffer are thus reduced.  The PMTs without concentrators can be used to study this background as well as help identify muons that cross the buffer and not the inner vessel.

All internal components of the detector were selected for low radioactivity~\cite{bib:ctf3}.  The predicted $\gamma$~background in the fiducial volume and in the range 250--800~keV is less than 0.5~counts/(day~$\cdot$~100~ton)~\cite{bib:ctf2,bib:ctf3}.

Besides being a powerful shield against external backgrounds ($\gamma$'s and neutrons from the rock), the Water Tank (WT) is equipped with~208~PMTs and acts as a \che muon detector (see Fig.~\ref{fig:detector}).  The muon flux, although reduced by a factor 10$^6$ by the~3800~m.w.e. depth of the Gran Sasso Laboratory, is still significant (1.1~muon~m$^{-2}$~h$^{-1}$) and an additional reduction (by about~$10^{4}$) is necessary.  See Ref.~\cite{bib:bxdetector} for a detailed description of the detector, the electronics, the trigger, and the calibration system.

Since the inner nylon vessel is in direct contact with the scintillator, extreme precautions were taken in the fabrication and assembly of the nylon vessels.  The nylon was selected for low radioactivity and extruded into film under carefully controlled conditions, then sealed in special bags to prevent dust and exposure to radon from the air.  The nylon vessels were built in a hermetically sealed Class-100 clean room.  A novel radon filter provided low-radon make-up air for the clean room to minimize surface deposit of the long-lived $^{210}$Pb radon daughter~\cite{bib:pocar}.  The two vessels were nested in the clean room, then shipped to the Gran Sasso Laboratory and installed in the SSS maintained as a Class-10,000 clean room.  For details on the nylon vessel fabrication and installation see Ref.~\cite{bib:nylon}.

After the nylon vessels were installed and inflated, they were purged to remove radioactive $^{39}$Ar and $^{85}$Kr.  Purging with high purity nitrogen was carried out in stages over a period of several weeks to permit Ar and Kr trapped in stagnation regions to diffuse into the main volume and be removed.  After thorough purging, the SSS and nylon vessels were filled with high purity water ($\approx 10^{-14}$~g/g of U/Th equivalent, 1~mBq/m$^3$ for $^{222}$Rn and $<$0.8~mBq/m$^{3}$ from $^{226}$Ra).  The water dissolved residual nylon monomers that caused haze and also removed metal ion impurities (such as $^{210}$Pb) from the surfaces of the vessels.

Underground purification of the scintillator by distillation, water extraction, nitrogen stripping and ultra-fine filtration were key for the Borexino detector to remove contaminants from dust (U, Th, K), from the air ($^{39}$Ar, $^{85}$Kr) and from cosmogenically produced isotopes (\ber). Extreme air tightness was maintained to prevent reintroduction of airborne impurities into the scintillator.  The scintillator solvent (PC) was distilled in-line during the detector filling at 80~mbar and at a temperature of 90--95\,$^{\circ}$C in a six-plate column with a high reflux ratio at a flow rate of~$\approx$800 l/hr.  Distilled PC was stripped in a 15~cm-diameter, 8~m-high packed column with specially prepared ultra-low Ar/Kr nitrogen (0.005~ppm Ar and 0.06~ppt Kr, see Ref.~\cite{bib:lakn}) at a reduced pressure of~$\approx$250~mbar and elevated temperature of  $\approx$35\,$^{\circ}$C.  In addition to removing radioactive contaminants the distillation of the PC was instrumental for obtaining  high optical clarity.

A concentrated master solution of the fluorescent component PPO in PC was prepared and purified in batch.  The master solution was filtered through 0.05~$\mu$m~filters, water extracted, vacuum distilled at 30~mbar and 200\,$^{\circ}$C in a falling film evaporator, and lastly nitrogen stripped.  The master solution was stored and metered to mix in-line with the PC solvent during the filling of the Borexino detector.  The filling of the Borexino detector was done under strict control of the flow rates to each of the three volumes of the detector, inner vessel, inner buffer and outer buffer.  The differential pressures, the liquid levels, and the strains in the supporting ropes were continuously monitored to sequence the flow to each volume to maintain uniform levels and differential pressures $<$0.5~mbar.  The shapes of the nylon vessels were maintained during the filling without introducing stresses that could cause creep and shape deformation.  The shapes of the vessels were also continually monitored with a specially developed internal array of cameras~\cite{bib:henning}.  The fluids were frequently sampled to test for optical quality, PPO and DMP concentrations.  The scintillator quality was verified  to satisfy stringent figures of merit.

The purification system and the filling modules were pre-cleaned following a precision cleaning procedure that employed high purity cleaning agents and deionized water.  The critical components of the system met the cleanliness level~30 by the standard MIL-STD-1246C for particulate specification for fluid contamination.  Once cleaned, all the equipment was sealed to avoid air contamination. To minimize air contamination in critical areas, the system was required to meet a stringent leak specification of $<$10$^{-8}$~bar~cm$^3$~s$^{-1}$.  Double seals with nitrogen blanketing between the seals were employed in several critical sections of the plants.  For more details concerning the purifications plants and operations see Ref.~\cite{bib:plants}.

\section{Data acquisition and event reconstruction}
\label{sec:data}

Monochromatic 862~keV neutrinos from \ber\ offer two signatures in Borexino.  The first is a recoil electron profile with a clear Compton edge at 665~keV that is the main spectral feature used in this work to separate the neutrino signal from the residual radioactive background.  The second possible signature, namely the~$\pm$3.5\% annual variation of the flux due to the Earth orbit eccentricity, cannot be considered yet because of the short running time so far compared to the Earth's orbital period.

The scattered electrons are detected by means of the scintillation light produced in the liquid scintillator.  The effective light yield was measured using the data by means of a spectral fit to the $^{14}$C background, which dominates the count rate below 160~keV.  The fit yields $\simeq$500~photoelectrons/MeV, after deconvolving the light quenching effect for low energy electrons \cite{bib:birks}.  This results is consistent with the known intrinsic light yield of the scintillator ($\approx$11,000~photons MeV), the geometrical coverage ($\approx$30\%), the PMT quantum and collection efficiency and the scintillator light propagation mechanisms that have been extensively studied in CTF and with small scale laboratory setups~\cite{bib:gemma,bib:ctf4}.  Light absorption from PPO and reemission with~80\%~probability and Rayleigh scattering from PC are the two dominant mechanisms influencing the light collection in Borexino and making the effective light yield very uniform for events produced within the fiducial volume.  The position of the $^{210}$Po charge peak and the $^{14}$C spectrum have been studied for events reconstructed within shells centered at various radii, obtaining the result that the effective light yield only decreases by~2\% going from events reconstructed in the detector center to events reconstructed at a radius of~3.5~m.

The scintillation light is collected by the~PMTs, and for each detected photon the arrival time and the charge are measured by means of a suitable analog and digital electronics chain.  As described in greater detail in Ref.~\cite{bib:bxdetector}, the Borexino main trigger fires when at least K$_{\rm thr}$~PMTs each detect at least one photoelectron within a time window of 60\,ns. The typical value was~K$_{\rm thr}$ = 30 in the data runs used in this paper, corresponding approximately to an energy threshold of  60 keV.  When a trigger occurs, the time and the charge of each PMT that has detected at least one photoelectron in a time gate of 7.2~$\mu$s is recorded.  The time is measured by a Time to Digital Converter (TDC) with a resolution of about 0.5~ns, while the charge (after integration and pulse shaping) is measured by means of an 8~bit Analog to Digital Converter (ADC).  This time resolution is smaller than the intrinsic time jitter of the PMTs which is 1.1~ns.

The trigger was designed to detect fast delayed coincidence with high efficiency. Events that are closer than 7 $\mu$s (such as \bipo\ and $^{85}$Kr delayed 
coincidences) are detected in a single trigger gate. Delayed coincidences separated by more than 7 $\mu$s are detected in two separate events with a dead time of 5 $\mu$s. The detection efficiency of \Bipo\ concidences exceeds 95\%.  

The readout sequence can also be activated by the outer detector by means of a suitable triggering system which fires when at least six PMTs detect light in a time window of 150~ns.  Regardless of the trigger type, both inner and outer detectors are always read.  For detector monitoring and calibration purposes, every two seconds, all inner detector PMTs are synchronously illuminated by a 394~nm laser pulse through a set of optical fibers that reach all PMTs. These laser triggers are used for precise time alignment and charge equalization and calibration of all channels. A similar system based on a 
set of LEDs is used for the outer detector.

The typical triggering rate during the runs analyzed in this paper was 15~Hz, including all trigger types.  This rate is largely dominated by very low energy $^{14}$C events.

Two independent offline analysis codes were employed to reconstruct the time, charge and position of each recorded event.  As a first step in the offline analysis, the codes identify in the recorded gate the clusters, that is, groups of time-correlated hits that belong to a unique physical event.  Generally, only one cluster is found in each event gate, but in the case of fast coincidences like \bipo\ and $^{85}$Kr-$^{\rm 85m}$Rb, or in the case of accidental pile up, more than one scintillation event may be recorded within the same trigger gate, yielding more than one reconstructed cluster from one event. Piled-up events resulting from the superposition of two different scintillation events within a few hundred ns are identified and rejected in the course of the analysis. The efficiency of the clustering algorithm was checked by means of random triggers and found to be nearly~100\% for events with more than~20~hits.

The position is determined using a photon time of flight method.  A probability density function (pdf) was developed using Monte~Carlo simulations 
and checked on the CTF and in Borexino. The resolution in the event position reconstruction is 13$\pm$2~cm in $x$ and $y$~coordinates, 
and 14$\pm$2~cm in $z$~at the relatively high \Bipo\ energies. The agreement of the above numbers demonstrates the near spherical 
symmetry of the detector.
The spatial resolution is expected to scale as $\sqrt{N}$ where $N$ is the number of detected photoelectrons \cite{bib:mccarty}. This was confirmed by 
determining the $^{14}$C spatial resolution to be 41$\pm$6~cm (1$\sigma$) at 140~keV, as expected.

\begin{center}
\begin{figure}
\centering\epsfig{file=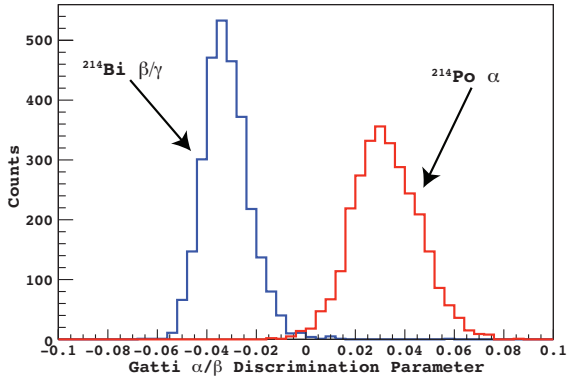, width=0.50\textwidth}
\caption{Distribution of the Gatti $\alpha/\beta$ discrimination variable for \Bipo\ events tagged by means of their fast 236~$\mu$s delayed coincidence.  The left solid blue curve is the distribution of the $\beta$~events, while the right solid red curve represents the $\alpha$~events.}
\label{fig:gatti}
\end{figure}
\end{center}

By measuring the position of each event we define a fiducial volume such that the residual external background mainly due to the PMT assembly, and in small part to the radioactivity of the nylon vessel, 
can be efficiently rejected.  The time profile of the hits of each cluster is also used to determine the event type. Particularly, muons crossing the buffer (if not tagged by the outer detector), $\alpha$~particles and $\beta$-like events can be identified by means of the mean time of the hits in the cluster, of the ratio between the number of hits in the tail and the total number of hits (tail-to-total ratio), and of several $\alpha/\beta$~discriminating procedures like the Gatti optimal filter~\cite{bib:gatti}.  An example of the $\alpha/\beta$~discrimination obtained by the Borexino detector is shown in Fig.~\ref{fig:gatti} for \Bipo\ events.

\begin{center}
\begin{figure}
\centering\epsfig{file=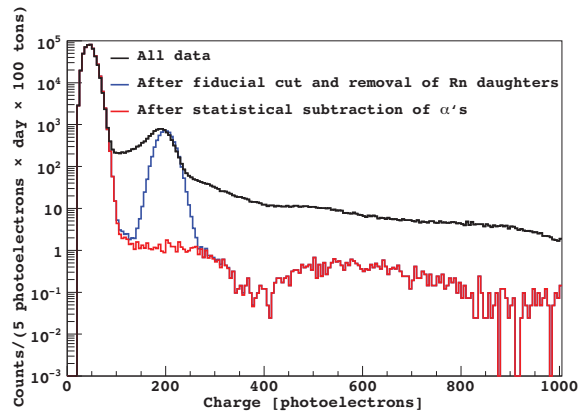, width=0.50\textwidth}
\caption{The raw photoelectron charge spectrum after the basic selection cuts (i--ii) (black), the spectrum after the subtraction of Rn daughters and muon-correlated activity, and after the application of the fiducial cut (iii--v) (blue), and with full $\alpha/\beta$ statistical subtraction of the $^{210}$Po $\alpha$ peak (red).  All curves scaled to the exposure of 100~day$~\cdot$~ton.  Cuts are described in the text.}
\label{fig:spectra}
\end{figure}
\end{center}

\section{Event selection and spectral fits}
\label{sec:fit}

The data used in this paper were collected for 47.4~live days between May and July~2007.  
Events are selected by means of the following cuts:

\vskip0.1in

\begin{enumerate}
\item The event must have a unique reconstructed cluster, in order to reject pile-up events and fast coincident events.  The efficiency of this cut is nearly~100\% because the very low triggering rate results in a negligible pile-up.
\item Events with a muon flag, i.e. those with some \che light in the water tank detector, are rejected.
\item After each muon that crosses the scintillator, all events (afterpulses and spurious events) within a time window of 2~ms are rejected.  The measured muon rate in Borexino (muons that cross the scintillator and buffer volume) is 0.055$\pm$0.002~s$^{-1}$.  The dead time introduced by this cut is negligible.
\item The Rn daughters occurring before the \Bipo\ delayed coincidences are eliminated by vetoing events up to three hours before a coincidence within a radial cut of 85~cm. The loss of fiducial exposure due to this vetoing process is ~0.7\%.
\item The events must be reconstructed within a spherical fiducial volume corresponding nominally to 100~ton in order to reject external~$\gamma$~background.  
Another volumetric cut ($z$$<$1.8~m) was applied in order to remove a small background from~$^{222}$Rn~daughters in the north pole of the inner vessel, 
resulting in a nominal fiducial mass of 87.9 t. The fiducial exposure in 47.4~days is 4136~day~$\cdot$~ton.
\end{enumerate}

\vskip0.1in

\begin{center}
\begin{figure}[!t]
\centering\epsfig{file=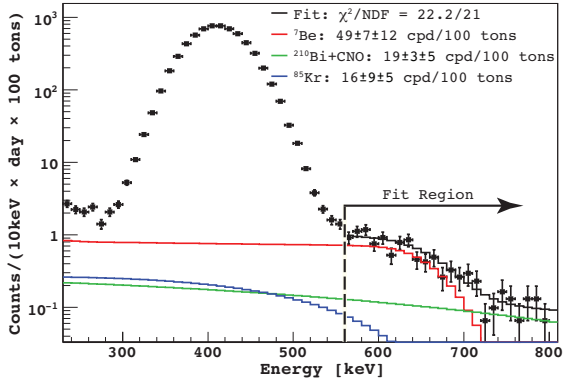, width=0.50\textwidth}
\caption{The fit to the \ber\ region without using $\alpha/\beta$ statistical subtraction.  The fit is done between 560 and 800~keV.}
\label{fig:echidnafit}
\end{figure}
\end{center}

Figure~\ref{fig:spectra} shows the effect of the selection cuts starting from the total raw count spectrum.  The solid black curve is the initial spectrum with only cuts (i-ii) applied.  At low energies, below 100~pe, the spectrum is dominated by $^{14}$C~$\beta$ decays intrinsic to the scintillator.  The peak at 200~pe is due to $\alpha$~particles from $^{210}$Po~decay, a daughter of $^{222}$Rn.  We find that $^{210}$Po is out of equilibrium with the other isotopes of the $^{222}$Rn sequence, as we see no evidence of a comparable amount of $^{210}$Bi and all precursors (see below).

The solid blue curve is the spectrum obtained after all of the above cuts (i-v) are applied.  The spectrum above 100~pe is significantly suppressed by the fiducial cut, by a factor $10^2 - 10^3$ with exception of the $^{210}$Po peak which is intrinsic to the scintillator.  Above the $^{210}$Po peak a clear shoulder is visible with an end point around 380~pe (approximately 750~keV), then the spectrum rises again mainly because of $^{11}$C~cosmogenic events.

Finally, the solid red curve is the spectrum that is obtained when the $^{210}$Po peak is statistically subtracted using the $\alpha/\beta$~discrimination algorithm (Gatti filter).  What remains in the region between 100 and 400~pe is dominated by \ber\ neutrinos.  In spite of the prominent $\alpha$~background, the sensitivity of the detector to \ber\ neutrinos is not compromised thanks to the excellent energy resolution and $\alpha/\beta$ discrimination.  Other spectral components ($^{85}$Kr, $^{39}$Ar, $^{214}$Pb, $^{210}$Bi) that may be present are not evident in this figure.

We assumed a linear relationship between energy and the number of detected photoelectrons in the 
limited region of the \ber\ neutrino events. The $^{14}$C spectrum was used to determine 
the conversion scale, accounting for the expected quenching of low energy electrons.

To measure the \ber\ neutrino signal we have followed two independent and complementary analysis approaches, 
which yielded consistent results. 
\begin{center}
\begin{figure}[!t]
\centering\epsfig{file=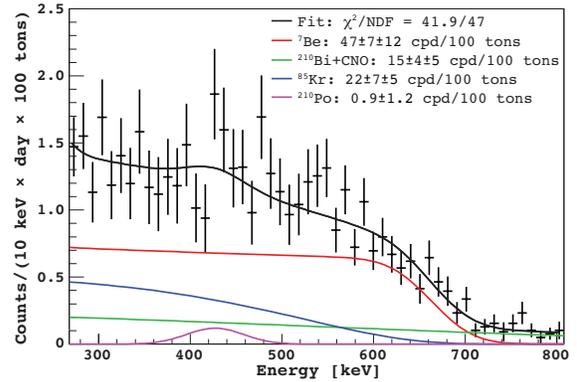, width=0.50\textwidth}
\caption{Spectral fit in the energy region from 270~keV up to 800~keV after $\alpha/\beta$ statistical subtraction of the $^{210}$Po~peak.}
\label{fig:mach4fit}
\end{figure}
\end{center}
In the first approach (see Fig.~\ref{fig:echidnafit}), events were selected with the additional requirement that the mean time of the hits belonging to the cluster with respect to the first hit of the cluster must be $<$100~ns.  This cut rejects residual muons that were not tagged by the outer detector and that interacted in the buffer (PC+DMP) regions.  Some $\alpha$~events are removed by this cut too.  Also, the exposure used for this analysis was 3058~day~$\cdot$~ton.
The shoulder in the spectrum between 560~keV and 800~keV was fitted taking into account the expected Compton like shape of the electron recoil spectrum from 0.862~MeV \ber\ neutrinos and the background from $^{85}$Kr and $^{210}$Bi $\beta$~decays.  The $^{85}$Kr component was constrained using data from the low energy region, between the $^{14}$C end-point and the $^{210}$Po peak.  In addition, the {\it pep} neutrino component was fixed to the expected SSM/LMA value and the contribution from CNO solar neutrinos and from $^{210}$Bi (which are not easily distinguished) were combined as a free parameter.  No other background components were included in the final fit because they were found not relevant.  As remarked above, $\gamma$~background in the fiducial volume and for the range 250--800~keV is lower than~0.5~counts/(day~$\cdot$~100~ton) and therefore is not relevant within the statistics accumulated~\cite{bib:ctf2}.  Moreover, the study of the fast coincidence decay of \Bipo\ (from $^{238}$U chain) and \bipo\ (from $^{232}$Th chain) yields, under the assumption of secular equilibrium, upper limits for $^{238}$U ($<$2~decays/(day~$\cdot$~100~ton), equivalent to 2$\times 10^{-17}$~g/g) and $^{232}$Th ($<$0.3~decays/(day~$\cdot$~100~ton), equivalent to 7$\times 10^{-18}$~g/g).  Consequently the contribution to the final spectrum of the background related to $^{238}$U and $^{232}$Th is negligible and was not included in the fit.

The shape of the $^{85}$Kr~$\beta$~spectrum and that of the \ber\ electron recoil spectrum are similar, and the two components are correlated in fits over the whole spectrum.  In the shoulder region the correlation is weaker, thus the \ber\ can be separated from $^{85}$Kr.  However, the spectrum cannot be fit with only $^{85}$Kr and this hypothesis is rejected at $>$5$\sigma$ level.

Our lack of knowledge of the real $^{85}$Kr content in the scintillator is one of the major sources of uncertainty, and will be further discussed below.  We searched for Kr through the rare decay sequence $^{85}$Kr$\rightarrow$$^{\rm 85m}$Rb+e$^-$+$\bar{\nu}_e$, $^{\rm 85m}$Rb$\rightarrow$$^{85}$Rb+$\gamma$ ($\tau$ = 1.5~$\mu$s, BR~0.43\%) that offers a delayed coincidence tag.  Only two candidates were found in the whole scintillator volume, with a corresponding upper limit of $<$35~counts/(day~$\cdot$100~ton) (90\%~C.L.) for dominant $^{85}$Kr $\beta$-decay branch.

In the second analysis approach (see Fig.~\ref{fig:mach4fit}) we performed a statistical subtraction of the $^{210}$Po~$\alpha$~peak by the following procedure.  For each energy bin, the distribution of the Gatti variable was fitted using two gaussian curves, one representing the $\beta$~population, the second one, the $\alpha$ population.  For each energy bin we fit and separated the two components, thus computing the relative amount of $\alpha$'s and $\beta$'s. In this way the $^{210}$Po was statistically removed, and the spectral fit could be performed in the whole energy range between 270 keV and 800 keV. The CNO and the $^{210}$Bi were again combined as a single fit parameter.  The fit also included a term for a residual $^{210}$Po peak, in order to account for possible inaccuracies in the statistical subtraction.

In both fits the statistical error is determined by three independent factors: a) the available statistics; b) the correlation of the fit result with the amount of $^{85}$Kr content; c) the correlation of the fit result with the exact value of light yield.
 
The first factor a) is trivially related to the amount of statistics collected so far; the second factor b) is related to the fact that we currently do not know the exact content of $^{85}$Kr in the scintillator, and no other method but the spectral fit itself could be applied so far. This uncertainty will be reduced by future calibration activities and also by the improved statistics, which will allow an independent measurement of the $^{85}$Kr content via the delayed coincidences.  We are also planning to determine the Kr content by direct measurement using mass spectrometry.  The third factor c) is due to the fact that our current determination of the light yield was obtained by fitting the upper half of the $^{14}$C~$\beta$~decay spectrum.  Although the position of the fitted \ber\ shoulder agrees very well with the expected value, the error in the light yield is still significant, and the exact \ber\ content is found to be correlated with the yield itself. The three errors combined in quadrature currently yield a statistical 
error of~15\%.

The main source of systematic error is the determination of the fiducial mass.  It was determined by rescaling the total number of background events from activities known to be uniformly distributed in the inner vessel to those found within the nominal fiducial volume, and by using the known amount of scintillator loaded into the inner nylon vessel (314.8~m$^3$, 278.3~ton).  In principle this procedure might allow the determination of the fiducial mass with fairly good precision.  However, the experimental distribution of the $^{14}$C is found to be non-uniform outside the fiducial volume, a fact that can be explained in part by a non-uniform detector response (light collection, trigger efficiency and other possible effects). Our best estimate of the systematic uncertainty currently associated with the definition of the fiducial mass is 25\%.  We expect to reduce the uncertainty significantly with an intensive calibration campaign.

When all effects are taken into account, our best value for the interaction rate of 0.862~MeV \ber\ neutrinos is: $47 \pm 7_{\rm stat} \pm 12_{\rm sys}$~counts/(day~$\cdot$~100~ton) in the full electron recoil spectrum.  The first quoted error is 1$\sigma$ statistical, while the second is the maximum systematic error due to the fiducial mass determination. 

The best value for the $^{85}$Kr rate given by the fit is $22 \pm 7 \pm 5$~counts/(day~$\cdot$~100~ton).  The fits also yield a contribution for the sum of $^{210}$Bi~$\beta$'s and the CNO signal of $15\pm4\pm5$~counts/(day~$\cdot$~100~ton).  The two components cannot be disentangled at this stage and the presence of other background sources to this number cannot be excluded.  Thus, the contribution due to the CNO component cannot be extracted from this number.

The excellent agreement of the observed spectrum with the expected spectral shape of a monenergetic 0.862 MeV neutrino strongly supports the conclusion 
that \ber\ solar neutrinos have been detected. We have not identified any background that  
can fit the spectrum in the region of the \ber\ $\nu$ signal. In particular,  
a potentially serious background in this energy region is $^{85}$Kr. However,  the data are  
highly inconsistent with the signal arising solely from $^{85}$Kr (the $\chi^2$ corresponds to a 
deviation of more than 5 $\sigma$). 

\section{Conclusions}
\label{sec:conclu}
We have measured the 0.862~MeV \ber\ component of solar neutrino spectrum in the Borexino detector.  The best value for the rate is $47 \pm 7_{\rm stat} \pm 12_{\rm sys}$~counts/(day~$\cdot$~100~ton).  The rate averaged over the earth orbit based on solar models and neutrino oscillations is expected to be 49$\pm$4~counts/(day~$\cdot$~100~ton) while the rate expected without oscillations is 75$\pm$4~counts/(day~$\cdot$~100~ton)~\cite{bib:fogli,bib:carlos}.
 
\section*{Acknowledgements}
We sincerely thank the funding agencies: INFN (Italy), NSF (USA), BMBF, DFG and MPG (Germany), Rosnauka (Russia) and we acknowledge the generous support of the Laboratori Nazionali del Gran Sasso.  We gratefully acknowledge the valuable contribution from former collaborators.  This paper is dedicated to the memory of  Cristina Arpesella, Martin Deutsch, Burkhard Freudiger, Andrei Martemianov and Sandro Vitale, and to John Bahcall, a friend and strong supporter of Borexino.

\end{document}